\documentclass[11pt,twoside]{article}
\usepackage{asp2006}

\usepackage{epsf}
\usepackage{psfig}
\usepackage{lscape}
\usepackage{graphicx}

\begin{document}

\title{Comparison of Two Solar Minima: Narrower Streamer Stalk Region and Conserved Open Magnetic Flux in the Region Outside of Streamer Stalk}   
\author{Liang Zhao and Len Fisk}   
\affil{Department of Atmospheric, Oceanic and Space
Sciences, University of Michigan, Ann Arbor, MI 48109-2143.}    

\begin{abstract} 
To explore the difference between the most two recent solar minima,
we analyze the in-situ  \emph{ACE} and \emph{ULYSSES} observations
and examine the distributions of the three types of solar wind
(streamer-stalk-associated wind, wind from outside the streamer
stalk that can be associated, in part, with coronal holes, and
interplanetary coronal mass ejections). We use the taxonomy provided
by \citet{Zhao09} to identify the three types of solar wind. We then
map the in-situ observations to the 2.5 solar radii surface. With
the aid of the potential-field-source-surface model (PFSS), we
calculate the normal distance from the solar wind "foot point" to
the local helisopheric current sheet on that surface. We find that
the source region of the streamer stalk wind is narrower ($15^\circ
\sim 20^\circ$) compared to the previous minimum ($\sim 40^\circ$).
The area outside the streamer stalk is accordingly larger, but the magnetic field strength is observed to be lower, with the result that the total amount of the magnetic open flux from the outside of streamer stalk region is conserved in the two successive solar minima. The implications of the conservation of open magnetic
flux for models of the behavior of the solar magnetic field are
discussed.
\end{abstract}

\section{Introduction}
There are three distinct types of solar wind identified by \citet{Zhao09}. First,
there is relatively high coronal electron temperature wind originating
from loops in the streamer stalk region \citep{WooMartin97}. Second,
there is solar wind from the outside of this region. This wind
includes coronal hole wind that has relatively low coronal electron
temperatures and high wind speeds, as well as slower
solar wind with lower coronal electron temperatures than the stream
stalk region. The third type of solar wind
is the transient interplanetary coronal mass ejections (ICMEs) which
are caused by the coronal mass ejections (CMEs)
\citep{RichCan95,ZurbuchenRich06,Burlaga02,Zurbuchen06}.

The streamer stalk region is the narrow region in the middle of the
streamer belt, which has the highest density fluctuations and the
lowest solar wind speeds \citep{Borrini81,Gosling81}.
\citet{WooMartin97} provide observational evidence that
the streamer stalks can be the coronal sources of the slow solar
wind. \citet{wang94} suggests that slow
solar wind originates from regions of rapidly expanding flux-tubes
located above small coronal holes and at the boundaries of the large
polar holes. Based on a comparison of solar remote and in-situ
observations, \citet{Liewer04} suggests that low-speed wind with
higher $O^{7+}/O^{6+}$  ratios may originate from open fields in or
near active regions. Fisk and collaborators
\citep[e.g.][]{Fisk99,Fisk03} suggest that
reconnection between open and closed field lines releases material to
form the solar wind. This model provides a
reasonable explanation for the differences between fast and slow
solar wind.

The distribution of the three types of wind varies with the solar
cycle. At solar minimum, the coronal holes concentrate at both poles
and high latitude coronal hole wind is observed \citep{Phillips95}. The heliospheric current sheet is flat and lies near the
equatorial plane, and the streamer belt stalk wind occurs in a band
around the current sheet \citep{Gosling97,Feldman81}. The ICME rate
is roughly proportional to the solar activity levels and therefore
is very low at solar minimum \citep{OwensCro06}. At solar maximum,
the current sheet tilts to high latitudes, and the streamer-stalk wind, which still occurs in a band around the current sheet, now can reach high latitudes. The polar coronal holes shrink, resulting in less coronal hole wind
in the heliosphere. The increasing rate of ICMEs can temporarily
enhance the open magnetic flux of the Sun. Subsequently, interchange
reconnection between the large ICMEs loops and the open field of
the Sun eliminates the increased magnetic flux
\citep{Gosling95,FiskSch01,Crooker02}. Further, there is no
compelling observational evidence to suggest that disconnection of
open magnetic flux occurs at the heliospheric current sheet
\citep{FiskSch01}. Hence, the expectation was, prior to this solar
minimum, the open magnetic flux would return to a constant
background level, as it had in previous minima
\citep{SvalgaardCliver07}.

In the current solar minimum, both the open magnetic flux and the mass
flux of the solar wind are reduced compared to any previous solar
minimum for which there are good space observations. In this paper,
we offer a possible explanation for the decrease in open magnetic
flux in the current minimum. We point out that the
streamer-belt-stalk-associated wind originates from a narrower
region in the current solar minimum compared to the previous one,
and thus the region outside the streamer belt stalk region is
larger. When we calculate the increase in area outside the stalk
region, we find it is equal and opposite to the decrease in open
magnetic flux, suggesting that the total magnetic flux in the region outside
the stalk region remains constant in each solar minimum. The
implication for the transport models of open flux developed by Fisk
and colleagues is then discussed.

\section{Observations at the Current Solar Minimum}
\subsection{The Changes of the Magnetic Field and Solar Wind Composition}

\begin{figure}[!ht]
\begin{center}
  \noindent\includegraphics[width=27pc]{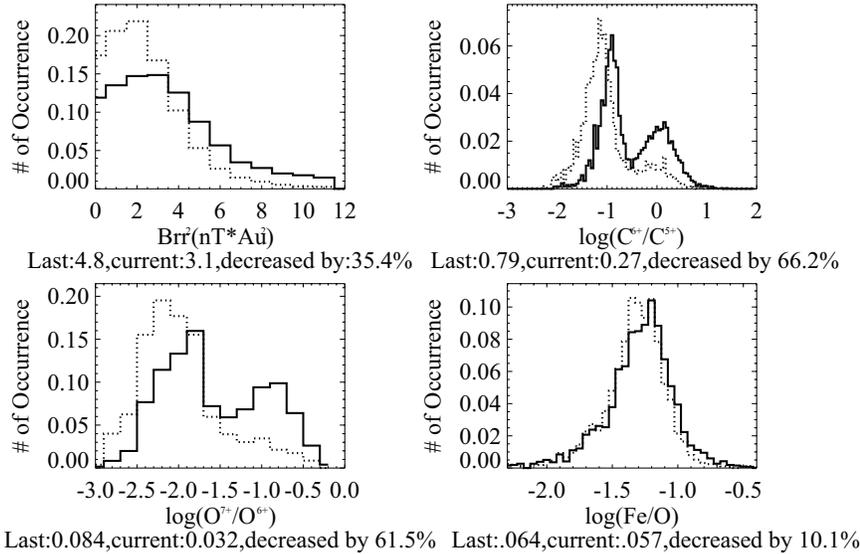}\\
  \caption{Histograms of $B_r r^2$ , $O^{7+}/O^{6+}$, $C^{6+}/C^{5+} $, and Fe/O at the previous (black) and the current (dotted) minimum from \emph{ULYSSES}.}
  \label{figure1}
  \end{center}
\end{figure}

The \emph{ULYSSES} 18-year mission started in 1991 \citep{Balogh92} and
terminated in 2009, providing sufficient data to compare the
different conditions in the two minima. Here we use the data from
the year of 1995.07-1998.2 (Carrington rotation 1892-1933) at last
minimum and 2005.83-2008.96 (Carrington rotation 2036-2077) at the
current minimum and compare the difference. As shown in Figure
\ref{figure1}, the radial component of the heliospheric magnetic
field, the so-called open magnetic flux of the Sun ($B_r r^2$), decreases by
35.4\%; $O^{7+}/O^{6+}$, $C^{6+}/C^{5+}$, and Fe/O all decrease by
61.5\%, 66.2\% and 10.1\%, respectively. Especially, when
considering the region outside of the streamer stalk (as identified
later), we find the open magnetic flux decreased by 30\%.

\begin{table}\footnotesize
  \centering
   \caption{In-situ signatures of three types of solar wind}\label{table1}
   \begin{tabular}{c c c l }
     \hline
       & Signature &      $V_{sw}$ Relationship & Criterion For \\
       \hline
     1 & $O^{7+}/O^{6+}$ & $O^{7+}/O^{6+}\geq 6.008 exp(-0.00578 V_{sw})$& ICMEs \\
     2 & $O^{7+}/O^{6+}$ & $0.145<O^{7+}/O^{6+}< 6.008 exp(-0.00578 V_{sw})$ & Streamer-stalk wind \\
     3 & $O^{7+}/O^{6+}$ & $O^{7+}/O^{6+}<0.145$ & Non-streamer-stalk wind \\
     \hline
   \end{tabular}
\end{table}

\subsection{The Narrower Streamer Belt Stalk Region}
\begin{figure}[!ht]
\begin{center}
  \noindent\includegraphics[width=25pc]{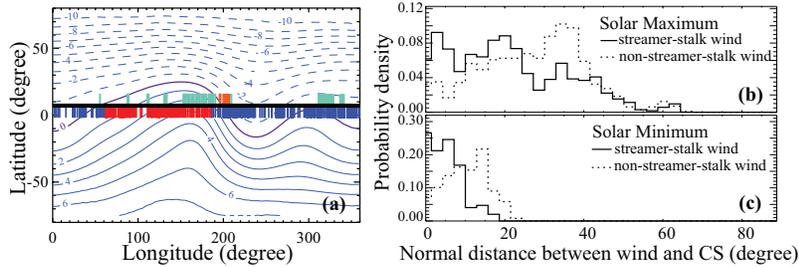}\\
  \caption{Right(a): Origin of three types of solar wind.
   Background contours shows the magnetic polarities from PFSS model:
   the dashed (solid) lines represent the inward (outward) magnetic field
   and the purple line is the current sheet. The black line in the middle
   of the color band is the trajectory of\emph{ ACE},
   the color bars above the black line indicate the two solar wind
   types (non-streamer-stalk wind in green and streamer-stalk wind in orange) and
    the color bars under the black line show observed magnetic
    polarities (inward in blue and outward in red). Left: (b)Probability densities
    of the normal distances from the source of streamer-stalk wind (solid line)
  and non-streamer-stalk wind (dotted line) to the local heliospheric
  current sheet on the 2.5 solar radii surface
   in the last solar maximum; (c) the current solar minimum.}
  \label{figure2}
  \end{center}
\end{figure}

We repeat the analysis of \citet{Zhao09} for \emph{ACE} observation to determine the three types of solar
wind: wind associated with the streamer belt stalk; wind from
outside this regions; and ICMEs. First, we identified the three-type
solar wind for the last solar maximum and the current solar minimum
using the criteria shown in Table \ref{table1}. Then, we map the
in-situ observations back to 2.5 solar radii to have a synoptic map
for each Carrington rotation
showing the distribution of the solar wind coronal sources, i.e.,
Figure \ref{figure2}a. As expected, \emph{ACE} observes
streamer-stalk wind (orange) when it crosses the current sheet. In Figure
\ref{figure2}a, the
polarities of magnetic field from observation match the PFSS model
very well. Based on these maps, we can calculate the normal distance
from the each of the "foot points" of the solar wind to the local
current sheet. Those normal distances are portions of great circle
arcs and can be expressed as an angle relative to the current sheet.

From Figure 2(b and c), in the current solar minimum, the width of
the streamer-stalk wind relative to the heliospheric current sheet
is only $\sim10^{\circ}$ on one side, or $20^{\circ}$ on both sides.
In the previous solar minimum, the width is
$40^{\circ}\sim50^{\circ}$ (Phillips 1995). Not surprisingly, the
normal distances from the streamer stalk wind foot points to the
heliospheric current sheet are more scattered at solar maximum than
minimum.

\subsection{The Conservation of the Total Magnetic Flux in the non-Streamer-Stalk Region}
We conclude above that in the current minimum the streamer stalk
region is narrower, and the area outside the streamer
stalk is larger, than the previous minimum. Consider, then, how the
total open magnetic flux contained in the region outside the
streamer stalk region varies between the two solar minima.

The total amount of the open magnetic flux is the product of the
area or solid angle ($\sigma$) occupied by the non-streamer-stalk region
and the magnetic strength ($B_rr^2$ ). From
Table \ref{table2}, the half-width of the streamer stalk region in the last
minimum is $\sim25^\circ$, and in this minimum it is reduced to
$\sim7.5^\circ-10^\circ$. If we set the solid angle covered by the
non-streamer-stalk region in last minimum as 1, then in the current
minimum this solid angle increases to 1.43. The magnetic field
strength outside the streamer stalk region is lower in the current
minimum by $\sim$70\%. Thus, the total amount of
open magnetic flux in the region outside of the streamer stalk region
remains the same in the two minima.
\begin{table}\footnotesize
  \centering
   \caption{Total amount of magnetic flux outside of streamer stalk}\label{table2}
   \begin{tabular}{c c c c c }
     \hline
       & Streamer half-width &Non-streamer-stalk &$B_r r^2$ &Total Magnetic    \\
       &   (degree)&         region solid angle & & flux         \\     \hline
     Last minimum    & 25 &1&1&1 \\
     Current minimum &7.5$\sim$10&$\sim$1.43&0.7&$\sim$1 \\
     \hline
   \end{tabular}
\end{table}

\section{A New Model for the Transport of Open Magnetic Flux on the Sun}
\begin{figure}[!ht]
\begin{center}
  \noindent\includegraphics[width=18pc]{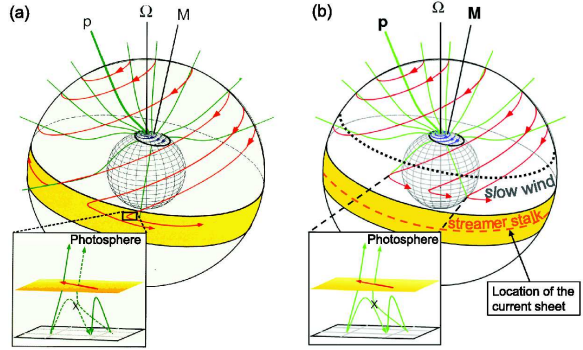}\\
  \caption{(a)An illustration of the motions of the magnetic field on
   the Sun in the frame corotating with the equatorial rotation
rate. The M-axis is the axis of symmetry for the expansion of the
magnetic field from a polar coronal hole. The $\Omega$-axis is the
solar rotation axis. P marks the open line (green) that
connects to the pole. The curves with arrows (red) are the
trajectories of the open field lines, and the yellow region is the
streamer stalk region. (b)The open lines reconnects and diffuses
outside the streamer stalk region. }
  \label{figure3}
  \end{center}
\end{figure}

Fisk and colleagues developed a model for the
global transport of open magnetic flux on the Sun, which is
illustrated in Figure \ref{figure3}a
\citep{Fisk96,Fisk99,Fisk05,FiskSch01}. Differential rotation drives the open flux across the polar coronal hole and then into closed field regions where open flux does not disconnect at the current sheet, but rather the flow patterns turn as shown. The process by which the magnetic field is transported through the closed field region is diffusion due to reconnection with loops. This model accounts for a number of features of the observed behavior of open magnetic flux; e.g., the slow solar wind appears to come from large coronal loops outside of coronal holes, and be released by reconnection \citep{Feldman05}.

This picture now needs to be revised, as shown in Figure
\ref{figure3}b. The open magnetic flux in regions outside the
streamer-stalk region is unable to penetrate into this region. Thus,
disconnection of this component of open flux, which must occur at
the heliospheric current sheet, is not possible. Rather, the turning
of the flow patterns of open flux must
occur outside the streamer stalk region, as shown. The total open
magnetic flux outside of the streamer-stalk region, which
cannot now disconnect, is conserved, as is observed to be the case.

\section{Conclusion Remarks}
There are several points worth emphasizing. The signature we use to
identify our streamer-belt-stalk wind is the $O^{7+}/O^{6+}$ ratio,
or the inferred coronal electron temperature, not the solar wind
speed, as was used in other studies. Moreover, it is important to
note that the streamer-stalk wind is not the entire slow speed solar
wind. Rather, it is only the very slow, high coronal electron
temperature wind, which we identify as originating from the streamer
stalk underlying the heliospheric current sheet. There is a broader
slow solar wind region \citep{Tokumaru09}, as constrained by the
black dotted line in Figure \ref{figure3}b.

The conservation of the total magnetic flux in the
non-streamer-stalk region during the two solar minima suggests that
the open magnetic field of the Sun in the current solar minimum is
behaving as it did in previous minima, the only difference being
the width of the streamer belt stalk region, which controls the
magnetic field strength in the region outside the streamer belt
stalk region.


\acknowledgements  This work was supported in part by NASA Headquarters under the NASA Earth and Space Science Fellowship Program-Grant NNX09AV13H, by the Heliophysics Theory Program, by NASA/JPL contract 1268016, and by NSF grant ATM 0632471.


\end{document}